\documentclass{aa}

\usepackage{psfig}

\def\mnras{MNRAS} 
\def\aap{A\&A}
\def\apj{ApJ}

\def\aj{AJ}
\def\apjs{ApJS}
\def\nat{Nat}

\let\a=\alpha
\let\l=\lambda
\def\ion#1#2{{\rm #1}\,{\sc #2}}

\usepackage{amssym}

\begin{document}



\title{A new deuterium abundance measurement from a
damped Ly$\a$ system at $z_{\rm abs} = 3.025$
\thanks{Based on public data released from UVES Commissioning at the 
VLT/Kueyen telescope, ESO, Paranal, Chile.}}

\author{S. D'Odorico \inst1, M. Dessauges-Zavadsky \inst1$^,$\inst2 and P. Molaro \inst3}

\institute{
European Southern Observatory, Karl-Schwarzschildstr. 2, D-85748 Garching bei M\" unchen, Germany
\and
Observatoire de Gen\`eve, CH-1290 Sauverny, Switzerland
\and
Osservatorio Astronomico di Trieste, Via G.B. Tiepolo 11, I-34131 Trieste, Italy
}


\mail{sdodoric@eso.org, mdessaug@eso.org,\\molaro@ts.astro.it}

\date{Received / Accepted}

\authorrunning{S. D'Odorico, M. Dessauges-Zavadsky and P. Molaro}

\titlerunning{Deuterium in a damped Ly$\a$ system}

\maketitle


\begin{abstract}

We present the first D/H measurement in a damped Ly$\a$ (hereafter DLA) system at $z_{\rm abs} = 
3.025$ towards QSO 0347--3819 obtained from the UVES-VLT spectra. The DLA absorber has a 
metallicity of [Zn/H] $\approx -1.25$ and a relatively simple velocity structure, with two 
dominating components detected in several metal lines. The hydrogen Lyman series can be followed 
down to Ly12 thanks to the high UV-Blue efficiency of UVES. The best fit of the Lyman series lines, 
and in particular of Ly$\epsilon$, Ly8, Ly10 and Ly12, relatively free of local contamination, is 
obtained when the \ion{D}{i} absorption is included in the two main components. The measured 
deuterium column density yields $\frac{\rm D}{\rm H} = (2.24\pm 0.67)\times 10^{-5}$ close to other 
low D/H values from Lyman limit systems. The corresponding values for the baryon to photon ratio 
and the baryon density derived from D/H are $\eta \approx  6\times 10^{-10}$ and 
$\Omega_b h^2 = 0.023$ respectively.
\end{abstract}



\section{Introduction}

Deuterium (D or $^2$H) is the only element entirely produced by nuclear reactions in the first 
minutes after Big Bang (Wagoner et al. 1967). The D yields are the most sensitive to the nuclear 
density at the nucleosynthesis epoch among the primordial light elements $^3$He, $^4$He and $^7$Li,
thus making the D abundance the most sensitive measurement of the baryon density in the universe 
(Wagoner 1973, Schramm \& Turner 1998).

Deuterium is currently measured in the local interstellar medium (ISM), (D/H)$_{\rm ISM} = (1.6\pm
0.1)\times 10^{-5}$ (Linsky et al. 1993), but since whenever it is cycled through stars it is 
completely burned away, extrapolation to the primordial D/H value requires a modeling of the 
Galactic chemical evolution. Direct D measurements of primordial material are thus of 
high interest. Adams (1976) suggested that almost primordial D could be measured in low metallicity 
absorption line systems in the spectra of distant quasars (QSOs). This was recently achieved for a 
few systems, but with conflicting results differing by almost an order of magnitude.

A few systems provide {\it high} D/H values (see e.g. Webb et al. (1997) who measure  $\approx 
2\times 10^{-4}$), while two other systems give a {\it low} abundance at 
$(3.39\pm 0.25)\times 10^{-5}$ (Burles \& Tytler 1998a,1998b). An even lower D/H estimation was 
obtained by Molaro et al. (1999), further discussed by Levshakov et al. (2000). Kirkman et al. (2000) 
measured an upper limit of $6.76\times 10^{-5}$. The handful of D detections found so far does not 
allow a firm conclusion. Different arguments favour a low primordial D/H ratio: the possible \ion{H}{i} 
contamination of the \ion{D}{i} absorption lines and, on the modeling side, the results by
Tosi et al. (1998) which predict for a variety of chemical evolution scenarios and to be consistent 
with the Galactic data a maximum decrease of the primordial D abundance by a factor of 3.

The paucity of suitable absorption systems for accurate D/H measurements is due to the fact that 
only absorption line systems with simple velocity structures and with intermediate \ion{H}{i} 
column densities allow the detection of the \ion{D}{i} lines. At too low column densities the 
\ion{D}{i} lines are too weak for detection, whereas at high column densities the lines are normally 
washed out by the saturation of the \ion{H}{i} line. We show here for the first time that in the 
latter case the deuterium signature can be successfully detected through the higher members of the 
Lyman series, when the target is a damped Ly$\a$ system ($\log N$(\ion{H}{i}) $\geq 20.35$) at high 
redshift. This approach was first suggested by Khersonsky et al. (1995).


\begin{figure}
\centerline{\psfig{figure=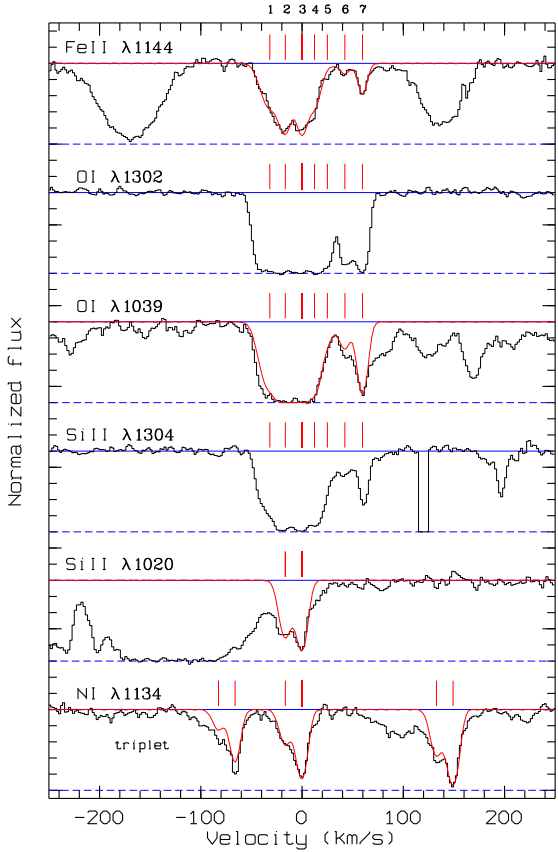,width=7.5cm,clip=}}
\caption{Absorption metal line profiles plotted against velocity for the DLA system at $z_{\rm 
abs} = 3.025$. The vertical scale goes from 0 to 1 for each plotted transition. The zero 
velocity is fixed at $z = 3.024856$. The vertical lines mark the positions of 7 components. The 
solid thin curve represents the best fit solution. The transitions \ion{O}{i}$\l$1302 and
\ion{Si}{ii}$\l$1304 are from the HIRES-Keck spectra published by Prochaska \& Wolfe (1999).}
\label{low-ion}
\end{figure}

  
\section{Data}

The spectra of QSO 0347--3819 (V=17.3, $z_{\rm em} = 3.23$) were obtained during the 
Commissioning of the Ultraviolet-Visual Echelle Spectrograph (UVES) on the VLT 8.2m Kueyen 
telescope at Paranal, in 1999. The instrument is described in D'Odorico et al. (2000). Two 
exposures of 4500 s each, covering the spectral range from 3650 to 4900 \AA\ and from 6700 to 
10000 \AA\ were obtained with a resolution of $\approx 6.9$ km s$^{-1}$ and $\approx 5.7$ km s$^{-1}$ 
respectively. The individual spectra were reduced using the UVES data reduction pipeline implemented 
in the ESO {\tt MIDAS} package. The final spectrum reaches a S/N varying from 20 to 35.

QSO 0347--3819 shows a damped Ly$\a$ system at $z_{\rm abs} = 3.025$ which has been studied in 
detail by Centurion et al. (1998), Ledoux et al. (1998) and from HIRES-Keck observations by 
Prochaska \& Wolfe (1999). The UVES observations however provide the first high-quality data in 
the UV ($\l < 4900$ \AA) and in the near-IR ($\l > 6700$ \AA). They allow the abundance measurements 
of new features such as O, P, Ar and Zn, in addition to the N, S, Si and Fe abundances measured in the 
previous studies. Full abundance analysis of the DLA system at $z_{\rm abs} = 3.025$ will be 
presented in a future paper. Here we focus on the deuterium detection and the D/H ratio measurement.


\begin{table}[t]
\begin{center}
\caption{Column densities} 
\label{HI-DI}
\begin{tabular}{l c l c c}
\hline
Comp  & $z_{\rm abs}$ & Ident      & $\log N$         & $b$           \\
      &               &            & [cm$^{-2}$]      & [km s$^{-1}$]       
\smallskip
\\     
\hline     
2...  & 3.024637      & \ion{H}{i} & 20.13 $\pm$ 0.08 & 21.5 $\pm$ 0.4 \\
      &               & \ion{D}{i} & 15.48 $\pm$ 0.08 & 14.1 $\pm$ 0.5 \\
3...  & 3.024856      & \ion{H}{i} & 20.43 $\pm$ 0.10 & 23.0 $\pm$ 2.0 \\
      &               & \ion{D}{i} & 15.78 $\pm$ 0.11 & 16.2 $\pm$ 3.0 \\
7...  & 3.025659      & \ion{H}{i} & 19.35 $\pm$ 0.05 & 14.7 $\pm$ 0.4 \\
\hline
\end{tabular}
\end{center}
\end{table}


\section{Analysis and results}

The absorption profiles of the DLA system at $z_{\rm abs} = 3.025$ are characterized by two 
dominating components (2 and 3) separated by about 20 km s$^{-1}$ with the red one slightly 
stronger than the blue as it can be seen from the non-saturated metal lines in Fig.~\ref{low-ion}. 
The strong and saturated metal lines reveal that additional material in smaller amount is present 
redwards the two main features (components 4, 5, 6 and 7). The region on the blue side of the main 
absorption components is sharp and relatively free from material with only one weak component 
(component 1) at about $-30$ km s$^{-1}$. In total seven components are needed to fit the metal lines 
absorption profiles, with two main components containing about 80\% of the total column density per 
transition.


\begin{figure}[b]
\centerline{\psfig{figure=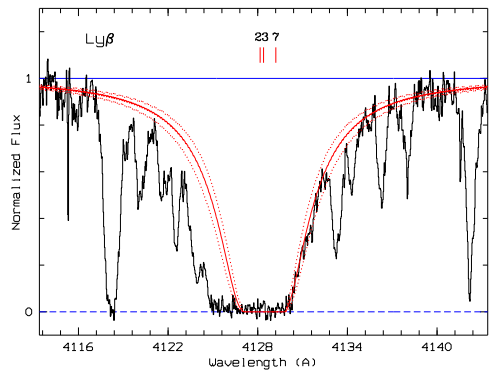,width=6.5cm,angle=-90,clip=}}
\caption{Damped Ly$\beta$ absorption profile. The vertical lines mark the positions of the three 
major hydrogen components (see text for details). The solid thin curve represents the best fit 
solution and the dotted thin curves delimit the range of uncertainty of the hydrogen column 
density.}
\label{Lyb}
\end{figure}


\begin{figure}
\centerline{\psfig{figure=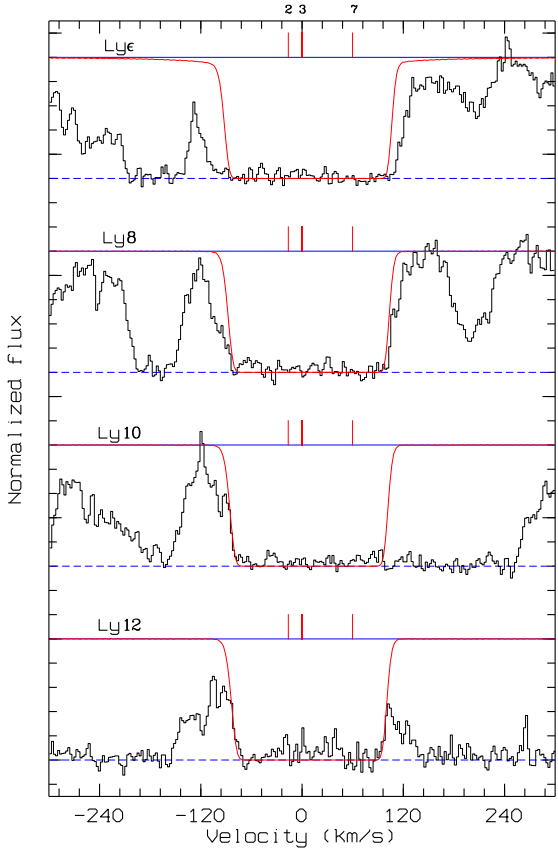,width=7.5cm,clip=}}
\caption{Ly$\epsilon$, Ly8, Ly10 and Ly12 profiles plotted against velocity. The vertical scale 
goes from 0 to 1 for each plotted transition. The zero velocity is fixed at $z = 3.024856$. The 
vertical lines mark the positions of the three hydrogen components. The solid thin curve 
represents the best fit solution.}
\label{Ly}
\end{figure}


The DLA system at $z_{\rm abs} = 3.025$ is a very good candidate for the deuterium analysis 
since it shows a relatively simple velocity structure dominated by two strong components, a low 
metallicity of [Zn/H]\footnote{[X/H] = $\log$(X/H) $- \log$(X/H)$_{\sun}$} $\approx -1.25$,
indicating that the measured D/H will be representative of primordial D/H and the hydrogen Lyman 
series lines can be followed down to Ly12.

We used a $\chi^2$ minimization routine {\tt fitlyman} (Fontana \& Ballester 1995) in {\tt MIDAS} 
to fit Voigt profiles to the observed absorption profiles, and obtain for each fitted absorption 
component the wavelength, the column density $N$, the Doppler parameter $b$ and the corresponding 
errors.

In the case of high hydrogen column densities like in DLA systems, we expect the neutral and 
low ionization metal lines to trace the H, therefore we model the Lyman series (Ly$\beta$ to Ly12) 
absorption profiles with three features corresponding to the two main components (2 and 3) and the 
reddest component (7) as determined from the metal lines (Fig.~\ref{low-ion}). The contribution of 
the weaker component 7 is required only to better constrain the fit on the red edge of the Lyman 
lines. The relative intensities of these three major hydrogen components were scaled with the metal 
lines assuming they have approximatively the same abundance ratios from component to component.


\begin{figure}
\centerline{\psfig{figure=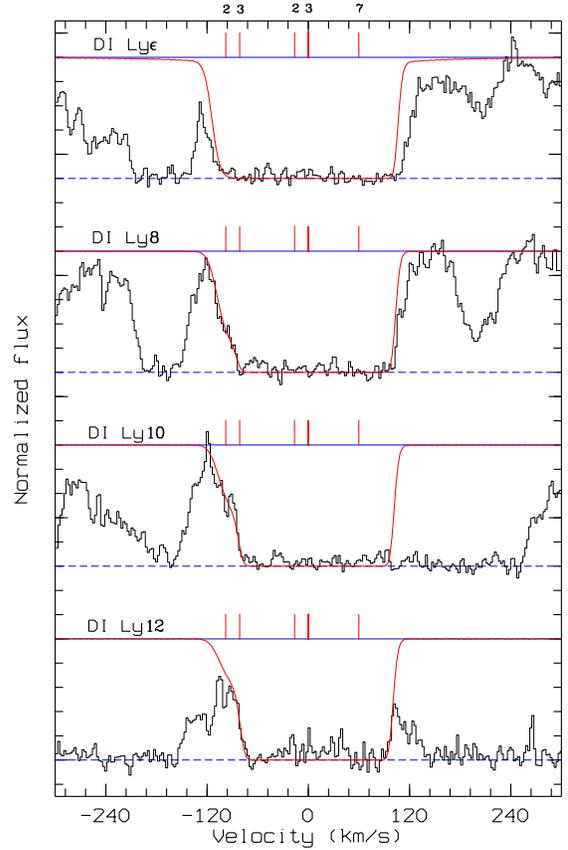,width=7.5cm,clip=}}
\caption{Same as Fig.~\ref{Ly}. The fit in this case is performed by considering the two main 
components for D and the three components for H (see text for details).}
\label{Ly-DI-2}
\end{figure}


Starting from this basis we obtained the final \ion{H}{i} column densities and  $b$-values (free 
parameters) by fitting simultaneously the lines of the Lyman series and by assuming the 
same redshift for the three H components as for the metal lines. The Ly$\beta$ absorption profile 
provides a very good constraint to the total \ion{H}{i} column density and the Ly8, Ly9, Ly10 and 
Ly12 profiles (which are the ones free from strong contamination) to the $b$-values. From the best 
fit (Fig.~\ref{Lyb}, Fig.~\ref{Ly}, Table~\ref{HI-DI}) we derived a total \ion{H}{i} column density 
of $\log N$(\ion{H}{i}) $= 20.63\pm 0.09$ in close agreement with the Pettini et al. (1994) value of 
$20.7\pm 0.1$. 

Adding other components to the hydrogen fit with low column densities as the ones observed in the 
stronger metal lines, component 1 at $-30$ km s$^{-1}$ and components 4, 5 and 6 at slightly higher
redshifts than the two main components, does not change significantly neither the total \ion{H}{i} 
column density of the two main components nor the fit on the blue and red wings of the Lyman 
lines. Fig.~\ref{Ly} however clearly shows that the fit with three hydrogen components 
systematically fails to reproduce the edge of the blue wing in the higher members of the hydrogen 
Lyman series, Ly$\epsilon$, Ly8, Ly10 and Ly12, at about $-82$ km s$^{-1}$ from the two H main 
components (2 and 3) which is the expected displacement of the corresponding D lines.

We then add to the model the deuterium by assuming its contribution only to the two main components 
(2 and 3). 
In the fitting procedure we assumed the same redshift for the two fitted components as for H and 
we left the column densities and the broadening parameters $b$ as free parameters.

The best fit performed over \ion{D}{i} Ly8 and Ly10 improves the $\chi^2$ of the hydrogen fit by a
factor of 3. It reproduces nicely the blue wing of the Ly8, Ly10 and Ly12 absorption profiles 
and is consistent with Ly$\epsilon$ (see Fig.~\ref{Ly-DI-2}). It gives a deuterium abundance of 
$$\frac{\rm D}{\rm H} = (2.24\pm 0.67)\times10^{-5} .$$
The error follows from the H and D fits and it includes the uncertainties in placing the continuum 
level and the errors on the $b$-values. The \ion{D}{i} column densities and the $b$-values of the 
two fitted components are given in Table~\ref{HI-DI}.

To check the stability of our result when the assumption that H and D mimic the metal structure is 
relaxed, we made a fit of the Ly$\beta$, Ly$\epsilon$, Ly8, Ly10 and Ly12 lines with three H and D 
components by keeping both their redshifts and column densities as free parameters. We obtained a 
satisfactory fit with shifts of the components with respect to the values given in Table~\ref{HI-DI}
of less than 4 km s$^{-1}$. The derived D/H ratio was comprised within our D/H error range. A fit 
with only one strong H component and the corresponding D component gave also a D/H ratio within our 
D/H error range.


\section{Discussion}

Three arguments support the interpretation of the absorption feature detected on the blue wing of 
some of the higher members of the \ion{H}{i} Lyman series, Ly$\epsilon$, Ly8, Ly10 and Ly12, as 
the \ion{D}{i} absorption with the same two main components as detected in the metal lines of the 
DLA system.

First, the absorption in the blue wing of the Lyman series lines could also be explained by a
hydrogen interloper associated with the damped Ly$\a$ system with a column density between 
$10^{15.7}$ and $10^{16.2}$, a $b$-value between 15 and 25 km s$^{-1}$ and placed between 
$-80$ and $-100$ km s$^{-1}$. However, from the density distribution of the Ly$\a$ clouds in the 
forest at $z\sim 3$ (see e.g. Kim et al. 2000), the probability to have such a cloud at that 
position is smaller than 1/1000. The lack of any metal component at this velocity in the strong and 
saturated lines of the DLA system (see Fig.~\ref{low-ion}) provides an additional evidence for 
discarding this possibility.

Secondly, the contamination on the blue wing of the Lyman series lines by different \ion{H}{i} 
interlopers at different redshifts which would mimic the same deuterium abundance as derived from 
the \ion{D}{i} Ly$\epsilon$, Ly8, Ly10 and Ly12 lines is even more unlikely.

Finally, the D and H column densities and broadening parameters $b$ resulting from the fits are 
consistent one relatively to other in the two fitted components (2 and 3): the derived D/H ratios are
the same and the $b$(\ion{D}{i})/$b$(\ion{H}{i}) ratios are close to what is expected in the 
thermally dominated case.

This measurement of D is the first made in a DLA system. It shows that the DLA systems with their low
metallicity ISM are a very promising class of absorbers for measurements of the D/H ratios 
at high redshifts, when it is possible to measure the higher members of the Lyman series. A 
systematic program of measurements using UVES data is under way.

The derived D/H ratio of $(2.24\pm 0.67)\times 10^{-5}$ is close to the low values obtained by 
Burles \& Tytler in Lyman limit systems (1998a,1998b) and it makes the claim of the primordial low 
D/H ratio more robust. Taken at face this ratio gives a baryon to photon ratio, $\eta = n_b/n_\gamma$, 
of $\approx 6.3\times 10^{-10}$ (Burles et al. 2000). This $\eta$ implies a helium abundance (in 
mass) of Y$_p$ = 0.2480 and a lithium abundance of $\frac{^7 \rm Li}{\rm H} \approx 
5\times 10^{-10}$ which are both larger than presently allowed by observations of He in 
extragalactic \ion{H}{ii} regions and of Li in halo stars (Izotov \& Thuan 1998, Bonifacio \& 
Molaro 1997). On the other hand an $\eta$ of $\approx 6.3\times 10^{-10}$ corresponds to a 
present-day baryon density of $\Omega_b h^2 = 0.023$\footnote{$H_0 = 100h$ km s$^{-1}$ Mpc$^{-1}$} 
which remains significantly lower than the $\Omega_b h^2 = 0.032^{+0.009}_{-0.008}$ derived from CMB 
anisotropy (Jaffe et al. 2000).
 

\begin{acknowledgements}
We are indebted to the UVES project team
for the high quality of the spectra obtained early in the operation of the instrument. We like to 
thank J.X. Prochaska for making available the profiles of the stronger metal lines and for 
comments on an earlier version of the manuscript.
\end{acknowledgements}




\begin{thebibliography}{}

\bibitem{} Adams T.F., 1976, \aap, 50, 461 
 
\bibitem{} Bonifacio P., Molaro P., 1997, \mnras, 285, 847

\bibitem{} Burles S., Tytler D., 1998a, \apj, 499, 699

\bibitem{} Burles S., Tytler D., 1998b, \apj, 507, 732
 
\bibitem{} Burles S., Nollett K.M., Turner M.S., 2000, astro-ph/0010171

 
\bibitem{} Centurion M., Bonifacio P., Molaro P., Vladilo G., 1998, \apj, 509, 620

\bibitem{} D'Odorico S., Cristiani S., Dekker H., et al., 2000, SPIE, 4005, 121

\bibitem{} Fontana A., Ballester P., 1995, Messenger, 80, 37
 


\bibitem{} Izotov Y.I., Thuan T.X., 1998, \apj, 500, 188

\bibitem{} Jaffe A. H., et al., 2000, astro-ph/0007333

\bibitem{} Khersonsky V.K., Briggs F.H., Turnshek D.A, 1995, PASP, 107, 570
 
\bibitem{} Kim T.-S., Cristiani S., D'Odorico S., 2000, \aap, submitted

\bibitem{} Kirkman D., Tytler D., Burles S., Lubin D., O'Meara J.M., 2000, \apj, 529, 655



\bibitem{} Ledoux C., Petitjean P., Bergeron J., Wampler E.J., Srianand R., 1998, \aap, 337, 51

\bibitem{} Levshakov S.A., Agafonova I.I., Kegel W.H., 2000, \aap, 355, L1

\bibitem{} Linsky J.L., et al., 1993, \apj , 402, 694 

 
\bibitem{} Molaro P., Bonifacio P., Centurion M., Vladilo G., 1999, \aap, 349, L13 


 
\bibitem{} Pettini M., Smith L.J., Hunstead R.W., King D.L., 1994, \apj, 426, 79

\bibitem{} Prochaska J.X., Wolfe A.M., 1999, \apjs, 121, 369
 

\bibitem{} Schramm D.N., Turner M.S., 1998, Rev. Modern Phys., 70, 303

\bibitem{} Tosi M., Steigman G., Matteucci F., Chiappini C., 1998, \apj, 498, 226



\bibitem{} Wagoner R.V., 1973, \aj, 179, 343

\bibitem{} Wagoner R.A., Fowler A., Hoyle F., 1967, \apj, 148, 3


\bibitem{} Webb J.K., Carswell R.F., Lanzetta K.M., Ferlet R., Lemoine M., Vidal-Madjar A., 
Bowen D.V., 1997, \nat, 388, 250


\end{thebibliography}
\end{document}